\def\cm2{cm$^{-2}$}

\def\c2{C~{\sc ii}}
\def\c4{C~{\sc iv}}
\def\fe2{Fe~{\sc ii}}
\def\fe3{Fe~{\sc iii}}
\def\mg1{Mg~{\sc i}}
\def\mg2{Mg~{\sc ii}}
\def\si2{Si~{\sc ii}}
\def\si4{Si~{\sc iv}}
\def\al2{Al~{\sc ii}}
\def\al3{Al~{\sc iii}}
\def\o1{O~{\sc i}}
\def\n1{N~{\sc i}}
\def\h1{H~{\sc i}}

\def\approxlt{\mathrel{\spose{\lower 3pt\hbox{$\sim$}}
        \raise 2.0pt\hbox{$<$}}}
\def\approxgt{\mathrel{\spose{\lower 3pt\hbox{$\sim$}}
        \raise 2.0pt\hbox{$>$}}}

\documentclass[article]{has40}
\usepackage{graphicx}
\usepackage{amssymb}
\tabletypesize{\normalsize}  

\def\plotone#1{\centering \leavevmode
\includegraphics[width=.95\columnwidth]{#1}}

\def\plottwo#1#2{\centering \leavevmode
\includegraphics[width=.45\columnwidth]{#1} \hfil
\includegraphics[width=.45\columnwidth]{#2}}

\def\plotone#1{\centering \leavevmode
\includegraphics[width=.95\columnwidth]{#1}}

\def\plottwo#1#2{\centering \leavevmode
\includegraphics[width=.45\columnwidth]{#1} \hfil
\includegraphics[width=.45\columnwidth]{#2}}

\shortauthors{Layden}
\shorttitle{Intrinsic Colors of RRc}

\begin{document}
\large    
\pagenumbering{arabic}
\setcounter{page}{102}

\title{Colors of c-type RR Lyrae Stars and Interstellar Reddening}

%
%
\author{{\noindent Andrew Layden{$^{\rm 1}$}, Tyler Anderson{$^{\rm 1,2}$} and Paul Husband{$^{\rm 1}$}\\
\\
{\it (1) Dept. of Physics \& Astronomy, Bowling Green State Univ., Bowling Green, OH, USA\\
(2) Dept. of Physics, Univ. of Notre Dame, Notre Dame, IN, USA} 
}
}

%
%
\email{laydena@bgsu.edu}


\begin{abstract}
RR Lyrae stars pulsating in the fundamental mode have long been used
to measure interstellar reddening, based on their observed uniformity
of $B-V$ color at minimum light after small corrections for
metallicity and period are applied.  However, little attention has
been paid to the first overtone pulsators (RRc or RR1).  We present
new $V-I$ observations of field RRc stars, supplemented with published
data from uncrowded RRc in globular clusters.  Preliminary results
indicate the RRc colors are correlated with period, but appear to be
independent of the stars' metallicity.  The scatter around the
period-color relation is slightly larger than a comparable relation
for RRab.  Thus, RRc can be useful indicators of line of sight
reddening toward old stellar systems, particularly when multiple stars
are available as in Oosterhoff II globular clusters and metal-poor
galaxies.
\end{abstract}

\section{Introduction}

The first figure of Horace Smith's book on RR Lyrae stars (RRL) demonstrates
the difference between fundamental mode pulsators (RRab or RR0) and
the first overtone stars (RRc or RR1): the latter have shorter
periods, typically $0.20 < P < 0.45$ days, and smaller amplitudes,
typically $\Delta V \approx 0.5$ mag (Smith 1995).  Preston (1964)
first noted that RRab have a constant color over the "minimum light"
phase interval of $0.5 < \phi < 0.8$, and Sturch (1966) utilized this
to develop a tight relation between intrinsic $B-V$ color, pulsation
period, and the photometric metallicity indicator $\delta(U-B)$.
Blanco (1992) improved this relation by adding more stars and casting
the metallicity in terms of the spectroscopic $\Delta S$ index of the
strength of the Ca II K line.  Walker (1990) provided a similar
relation using [Fe/H] to indicate metallicity.

Since then, more and more RRL photometry has been done using
red-sensitive CCDs in the $VI$ bands.  Mateo et al. (1995) showed that
$V-I$ colors can also be used to determine interstellar reddening, and
that the metallicity sensitivity is weaker than for $B-V$ based on
their sample of a dozen stars drawn from the literature.  Day et al.
(2002) and Guldenschuh et al. (2005) refined and extended this
relation for RRab using new observations of six southern stars.
Kunder et al. (2010) extended the relation to $V-R$ colors.

Despite this extensive development for RRab stars, little work has
been done to calibrate a reddening relation for RRc.  One example is
the relation presented in McNamara (2011),
\begin{equation}
(\langle B \rangle - \langle V \rangle)_0 = 0.048 + 0.501P,
\end{equation}
where the angled brackets indicate the intensity mean magnitude over
all phases.  McNamara derived this relation for the RRc in M3, and
extended it to other metallicities using a color correction based on
model atmospheres.  Our goal in this study is to provide reliable
calibrations for $B-V$ and $V-I$ colors based on a larger sample of
RRc stars spanning a wide range in metallicity and including stars in
both globular clusters and the Galactic field.  


\section{Data}
Over the last decade, we have acquired numerous $VI$ images of RRL
using the 0.5-m Cassegrain telescope and Apogee Ap6e CCD camera
located on the BGSU campus.  Here, we present data for 11 RRc.  For
each star, all-sky photometry was obtained using the WIYN 0.9-m
telescope at Kitt Peak in order to calibrate the $VI$ magnitudes of
about ten surrounding comparison stars to the Landolt (1992) scale,
and thereby enable the transformation of the differential photometry
of the RRc taken at BGSU onto the standard $VI$ system.

Several southern RRc were observed using the PROMPT 0.4-m
telescopes\footnote{Operated by the University of North Carolina at
Chapel Hill; we thank Dan Reichart and his team for providing access
to these facilities.} on Cerro Tololo, Chile.  Differential photometry
was derived from most images, and calibrated to the standard $VI$
system using Landolt (1992) standards taken on one photometric night.

Additional $BVI$ data were gleaned from published studies of RRc in
globular clusters.  For consistency with the field star data, we
preferentially selected studies with photometry tied to Landolt (1992)
or Stetson (2000) standard stars.  Table 1 shows the clusters we have
currently utilized, including the metallicity and reddening from the
Harris (1996) compilation, the Oosterhoff type, the number of RRc
stars in the cited reference, the filters employed, the standard stars
used to calibrate the photometry, the method of photometry, and the
reference to the study from which the photometry was taken.  In
most cases, we were able to retrieve the original time series
photometry, so we can plot the light curves in all available passbands
and perform our own statistical analysis on the data.  For many of
these clusters, more recent photometry is available, but it tends to
be of variables in the inner, crowded region of the cluster.  This
summer, we are in the process of adding 3-4 new clusters to the
analysis.

\begin{flushleft}
\begin{deluxetable*}{lccccccll}
\tabletypesize{\normalsize}
\tablecaption{Globular Clusters}
\tablewidth{0pt}
\tablehead{ \\ \colhead{NGC}   & \colhead{[Fe/H]} &
       \colhead{$E(B-V)$} & \colhead{Oo} & \colhead{$N_{RRc}$} &
  \colhead{Filters} & \colhead{Stds{\tablenotemark{a}}}  
  & \colhead{Method{\tablenotemark{b}}}  & \colhead{Reference} \\
}
\startdata
1851    & --1.18 & 0.02 &  I &  7 & $BVI$ & L92 & DAO & Walker (1998) \\
5904=M5 & --1.29 & 0.03 &  I & 14 &  $VI$ & L92 & DAO & Reid (1996) \\
5272=M3 & --1.50 & 0.01 &  I & 47 & $BVI$ & S00 & ISIS & Benk\H{o} et al. (2006) \\
4171    & --1.80 & 0.02 & I? & 10 & $BVI$ & S00 & DAO & Stetson et al. (2005) \\
4590=M68 & --2.23 & 0.05 & II & 16 & $BVI$ & L92 & DAO & Walker (1994) \\
 \enddata
\tablenotetext{a}{Standard stars from Landolt (1992) or Stetson (2000).}
\tablenotetext{b}{Photometry derived from DAOPHOT (Stetson 1987) or ISIS (Alard, 2000).}
\end{deluxetable*}
\end{flushleft}

\begin{figure*}
\centering
\plottwo{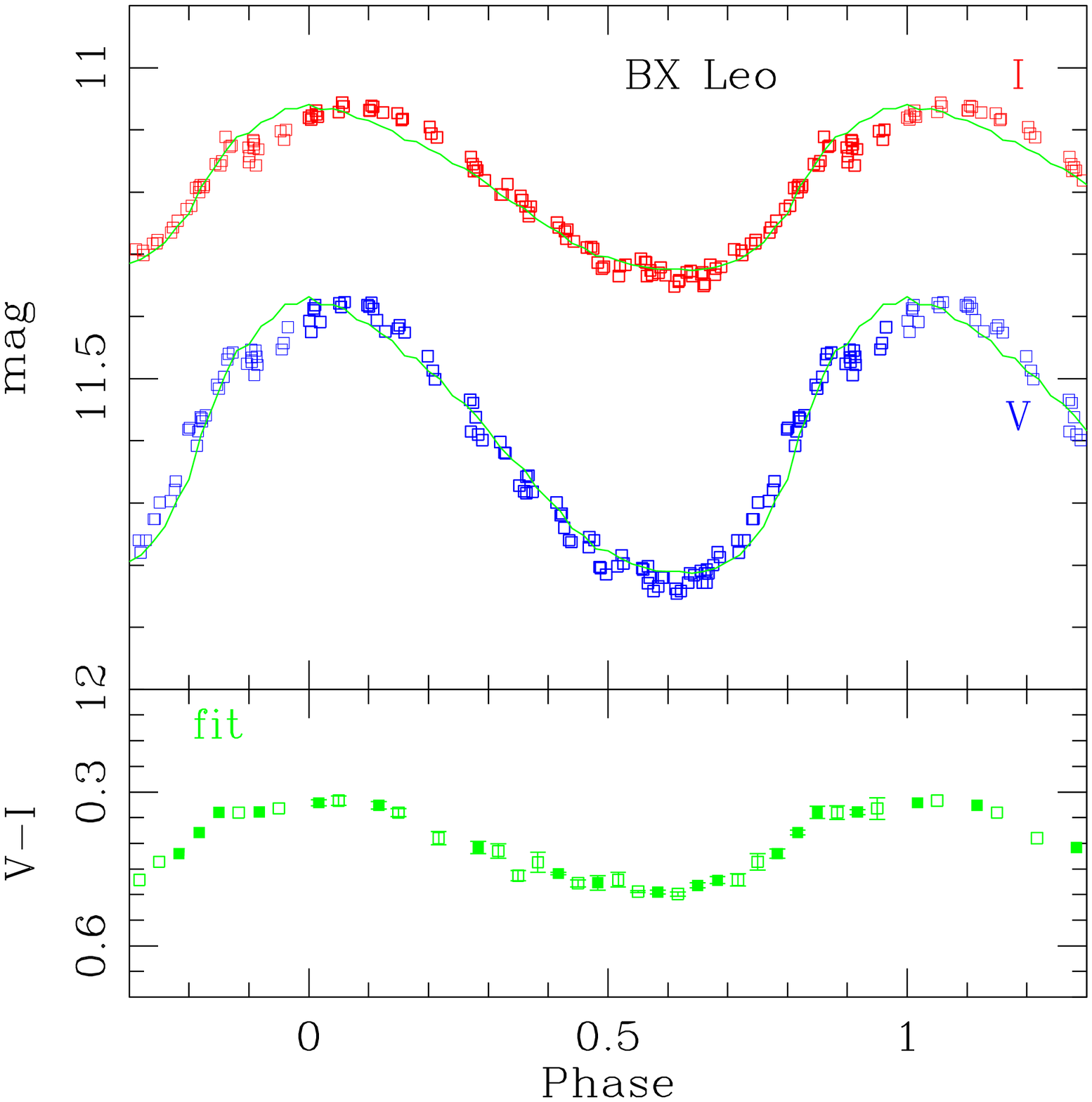}{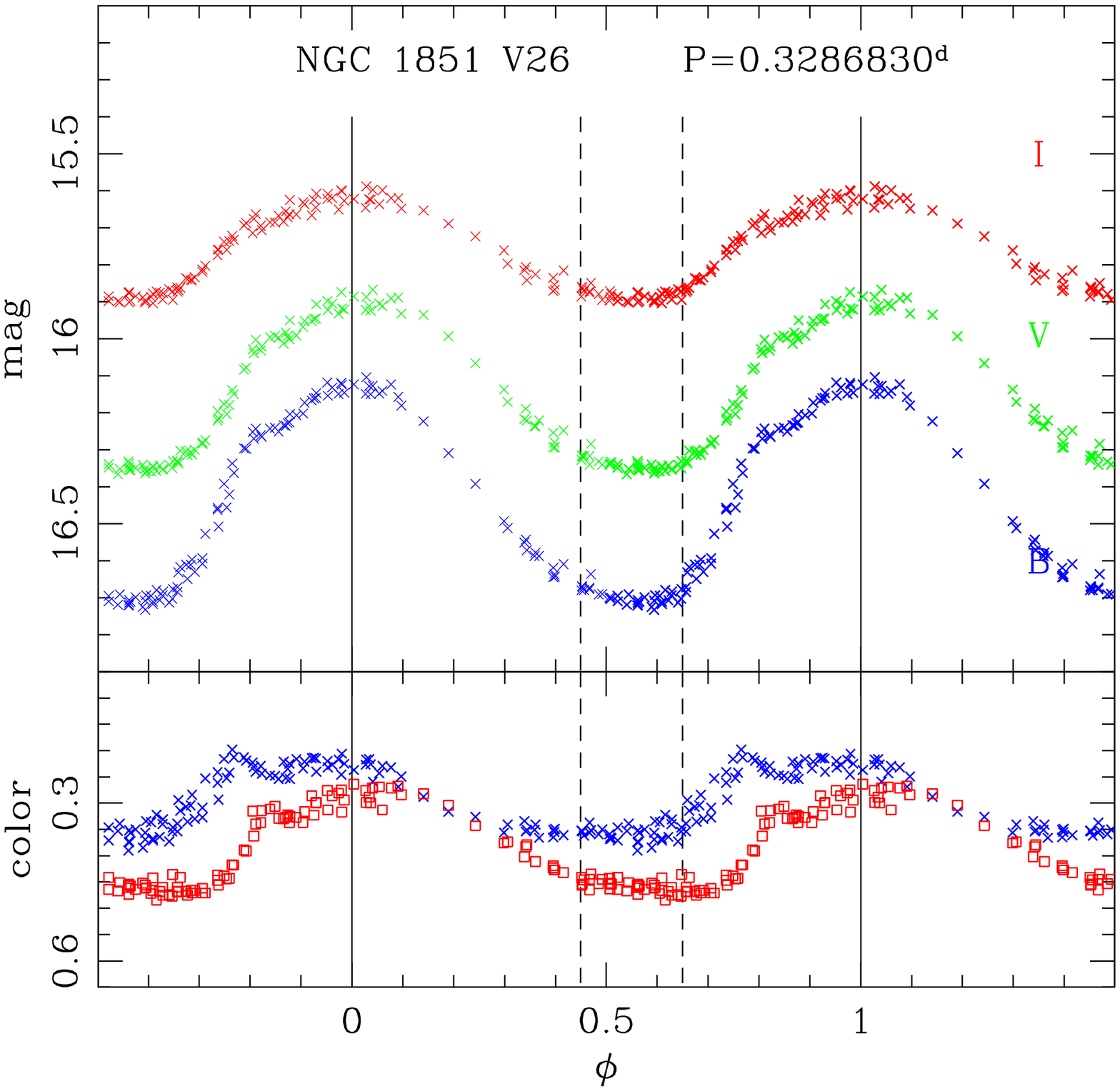}
\vskip0pt
\caption{The light and color curves for the field star BX Leo observed
at BGSU $(left)$, and the globular cluster star V26 in NGC 1851 from
Walker (1998), which has $C_C = 1 ~(right)$.  The red and blue color
curves for V26 show $V-I$ and $B-V$, respectively.}
\label{fig_lcs}
\end{figure*}

Light curves for two typical stars are shown in the top panels of
Figure \ref{fig_lcs}.  For stars with non-contemporaneous $BVI$
observations we created color curves by creating a series of twenty
phase bins, each 0.05 phase units wide, and calculating the
magnitude-mean $B$, $V$, and $I$ in each bin, and then calculating
$B-V$ and $V-I$ in each bin.  Results are shown in the bottom panels
of Figure \ref{fig_lcs}.

Because photometric crowding and blending can bias photometry in
globular clusters, we visually inspected the environs of each variable
star on a wide-field, mosaicked CCD image provided with the Stetson
(2000) standards,\footnote{See 
\tt{http://www1.cadc-ccda.hia-iha.nrc-cnrc.gc.ca/community/STETSON/}.} 
using finder charts or equatorial coordinates to identify the
variable.  We assigned the star a ``crowding class'' ($C_C$) on a 0-4
scale, where 0 indicates an isolated star comparable to most field
stars, and 4 indicates a badly blended image.\footnote{See 
\tt{http://physics.bgsu.edu/\~layden/BGSU\_Observatory/CrowdingClass.pdf}.}
While admittedly more subjective than the separation index of Stetson,
Bruntt \& Grundahl (2003), we do not have access to the original CCD
images or DAOPOT photometry from which to calculate the separation
index.  The crowding class has the advantage that it includes stars
not in the DAOPHOT star list, such as badly saturated stars, and it
considers the level of unresolved background light.  It is also a good
training exercise for acquainting new students with astronomical image
analysis and star field recognition.

\begin{figure*}
\centering
\plotone{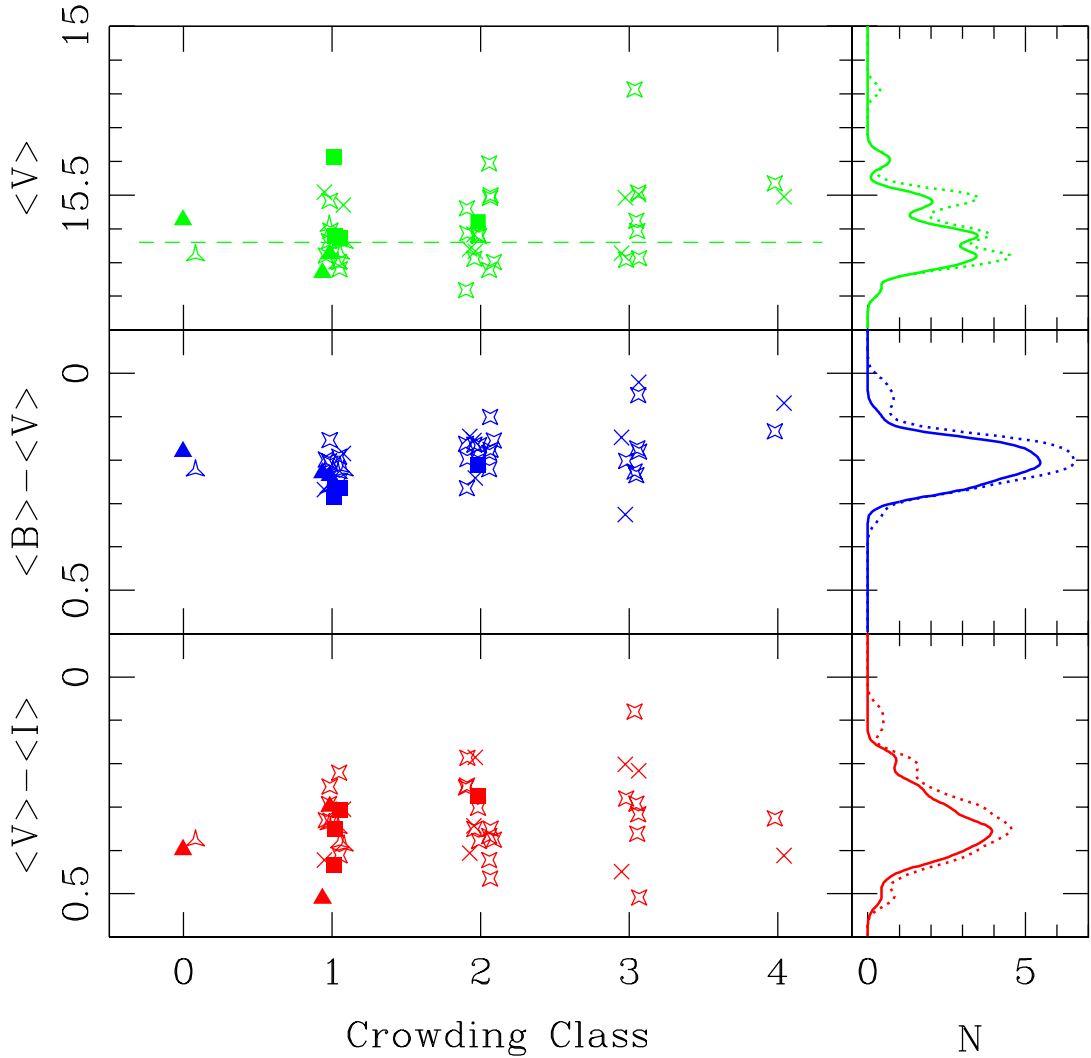}
\vskip0pt
\caption{The left panels show the intensity-mean $V$ magnitude (top),
$B-V$ color (middle), and $V-I$ color (bottom) as functions of
crowding class for the RRc stars in M3.  The symbols indicate light
curves with little scatter (solid), some scatter (open), and large
scatter (crosses), with 3-point and 4-point symbols indicating data
from the Schmidt and RCC telescopes, respectively (Benk\H{o} et
al. 2006).  The right panels show generalized histograms of these
distributions.}
\label{fig_cc}
\end{figure*}

The RRL-rich cluster M3 provides an opportunity to check how
photometry may be biased by stellar crowding, though this cluster is
far from the most crowded in our list.  In Figure \ref{fig_cc} we show
the intensity-mean magnitudes and colors of the RRc in M3 plotted
against their crowding class.  The right-side panels show generalized
histograms (replacing each binned point in a normal histogram with a
unit Gaussian having $\sigma = 0.02$ mag) of the stars having $0 \leq
C_C \leq 2$ (solid curves) and for all stars (dotted curves).  In the
top panel, we would expect to see more crowded stars show a tendency
to be brighter and span a wider range of magnitudes as neighbor stars
blend and merge with the target variable.  We see some of this
behavior, but there is also vertical scatter at every crowding class,
perhaps due to binary companions.
The colors show more affect of crowding: in both cases the scatter
increases with crowding class, though some of the scatter seen in
$\langle V \rangle - \langle I \rangle$ at all crowding classes is
attributable to the strong period-dependence described below.  Also
shown in the figure is the fact that the point-to-point scatter in a
star's light curve tends to increase with crowding class, suggesting
that the effects of crowding vary from image to image depending on
factors like seeing, sky background, etc.  From this analysis, we
conclude that the photometry of stars with $0 \leq C_C \leq 2$ is
generally trustworthy, while that of stars with $C_C = 3$ or 4 becomes
increasingly less so.
Since our goal is to get the best possible mean colors, rather than a
complete sample, and many clusters have abundant RRc, we can reject
stars with $C_C \geq 3$.

\section{Analysis}

The color of an RRL can be calculated in many ways.  Past studies of
RRab stars have focused on the mean magnitude-averaged color during
minimum light, defined as the interval between phases of 0.5 and 0.8
(Sturch 1966), where the color is approximately constant.  The
advantage of this definition is that when the number of data points is
small, as in a program to find RRL in a distant galaxy (e.g., M33,
Sarajedini et al. 2006) or heavily reddened cluster (e.g., NGC
6441, Pritzl et al. 2003), a single observation in this phase
interval may result in a better reddening estimate than the average of
all the data points over a light curve with sparse or gappy phase
coverage.
A disadvantage of minimum light colors, noted by McNamara (2011), is
that many extant studies (including those listed in Table 1) do not
tabulate minimum-light colors, necessitating recovery of the original
time-series data and its analysis.

The aim of our present study is to maximize the utility of our $V-I$
and $B-V$ color data by expressing the color in terms of all three of
the current definitions, and to explore which definition might produce
the smallest scatter and hence serve as the best reddening indicator.
We calculate minimum light colors from the original time-series, but
adopt from each study their tabulated intensity-mean colors, $\langle
V \rangle - \langle I \rangle$ (where the $\langle \rangle$ notation
indicates an intensity-mean magnitude taken over all phases), and the
magnitude-mean colors, ${V-I}_{mag}$ taken over all phases (our
calculated values usually match the tabulated ones to several 0.001
mag).

\begin{figure*}
\centering
\plotone{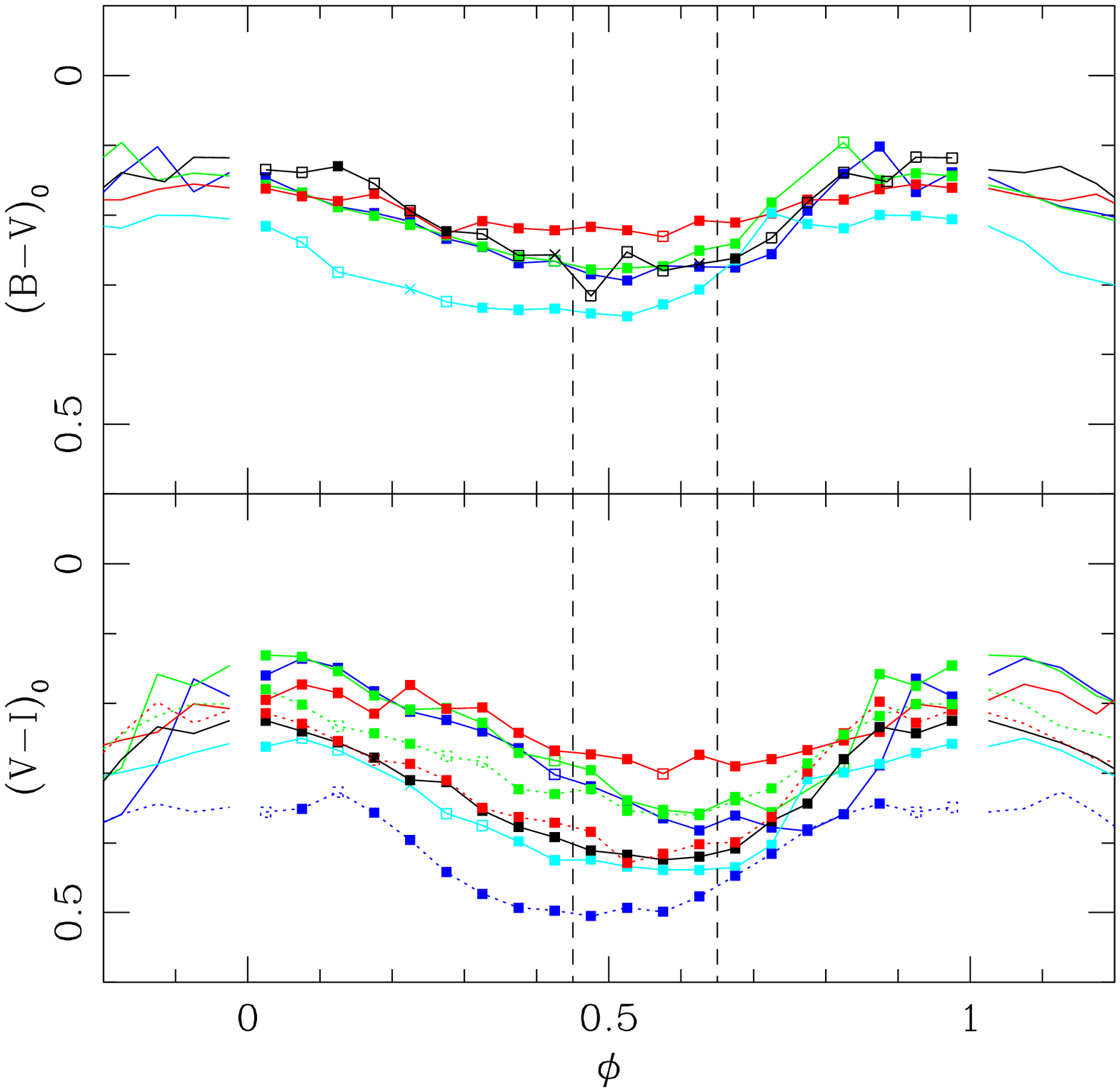}
\vskip0pt
\caption{Color curves for a selection of RRc in globular clusters.
The phase points 0.45 and 0.65 are marked.}
\label{fig_min1}
\end{figure*}

\begin{figure*}
\centering
\plotone{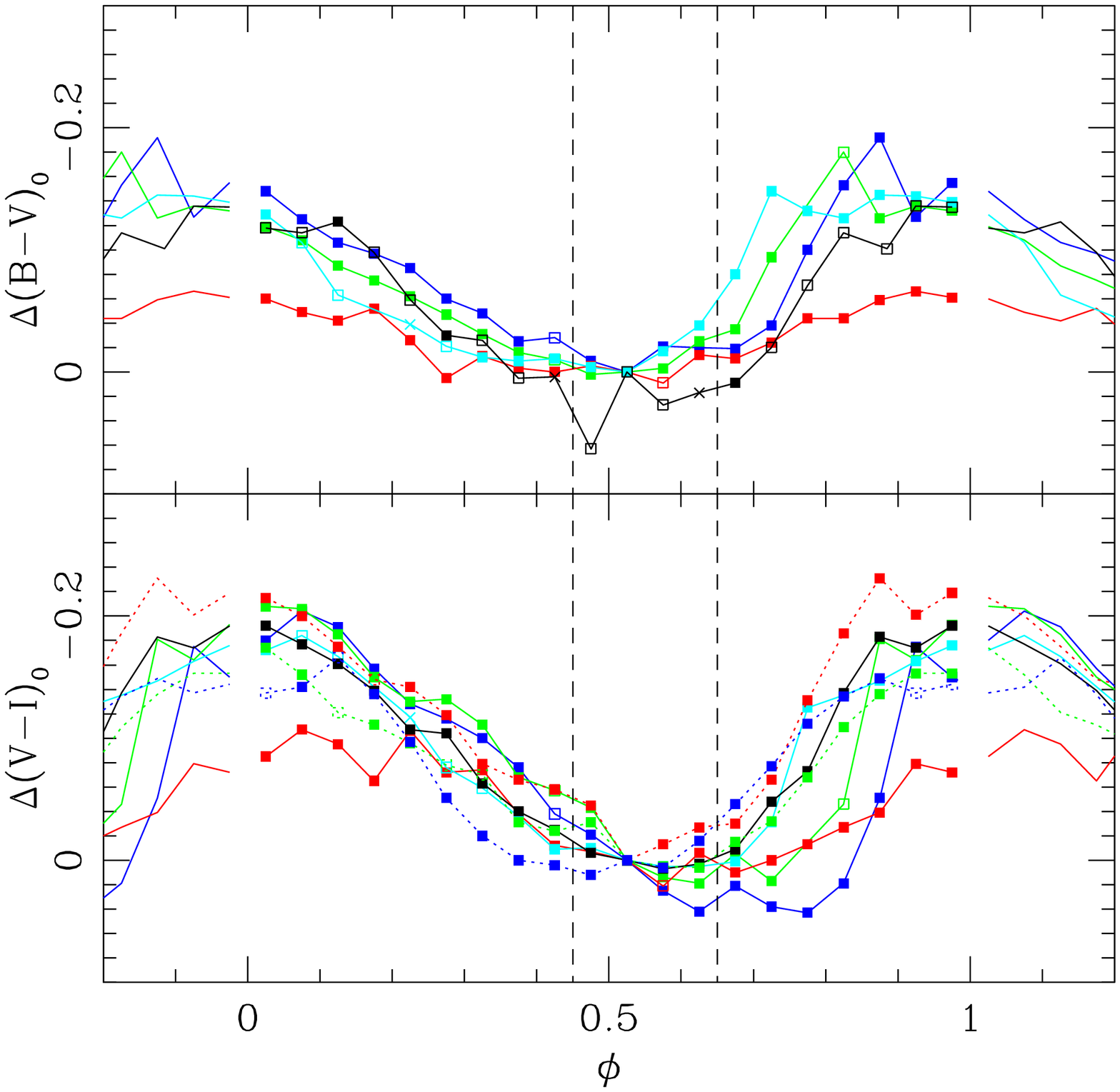}
\vskip0pt
\caption{The color curves from Figure \ref{fig_min1} shifted vertically to register at $\phi = 0.525$.}
\label{fig_min2}
\end{figure*}

Our first task is to determine whether RRc stars have a minimum light
phase interval in which the color is roughly constant.  We selected
about four stars from each cluster that have low crowding class, light
curves with little scatter, and which span a wide range in period.
Figure \ref{fig_min1} shows the color curves of these stars after
correction for the small reddenings listed in Table 1.  The goal was
to get a representative sample of light curve shapes and periods at
each cluster's metallicity.  It is clear from Figure \ref{fig_min1}
that the phase interval used for RRab, $0.5 \leq \phi \leq 0.8$, is
too large for the more sinusoidal light curve shapes of RRc.  However,
star-to-star variations in color due to period (see below) tend
to confuse and mask the behavior of the RRc near minimum light.
Following the method used by Sturch (1966), we vertically shift each
light curve so they match at a common phase point, in our case we
chose $\phi = 0.525$ as shown in Figure \ref{fig_min2}.  For $B-V$, it
seems that many stars have a fairly constant color over the phase
interval $0.45 \leq \phi \leq 0.65$, but in $V-I$ the stars have a
wider range of color curve shapes so that some stars become
bluer during this phase interval while others become redder.  As
the summer progresses and we add more clusters to Figures
\ref{fig_min1} and \ref{fig_min2}, we will perform a quantitative
analysis like Sturch (1966) to define the best minimum light interval
for RRc stars, but for now it appears that $0.45 \leq \phi \leq 0.65$
provides a good working definition.

\section{Analysis of Colors}

\begin{figure*}
\centering
\plotone{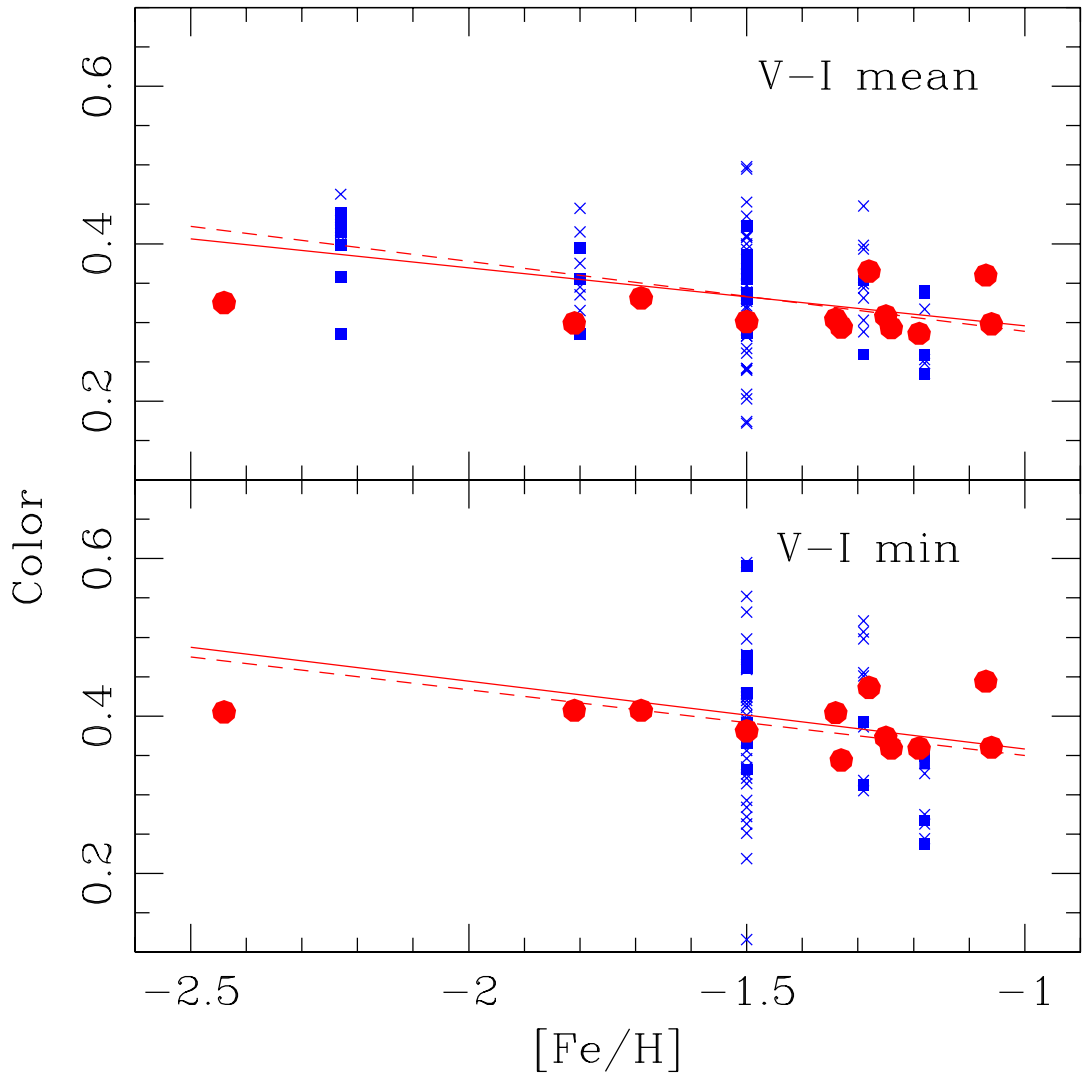}
\vskip0pt
\caption{Regressions of $(\langle V \rangle - \langle I \rangle)_0$
and $(V-I)_{min,0}$ versus [Fe/H].  Field and cluster stars are shown
in red and blue, respectively, with stars having high crowding class
marked with crosses.  The dashed and solid curves use all the data,
and the stars with low crowding class, respectively.}
\label{fig_fit1}
\end{figure*}

\begin{figure*}
\centering
\plotone{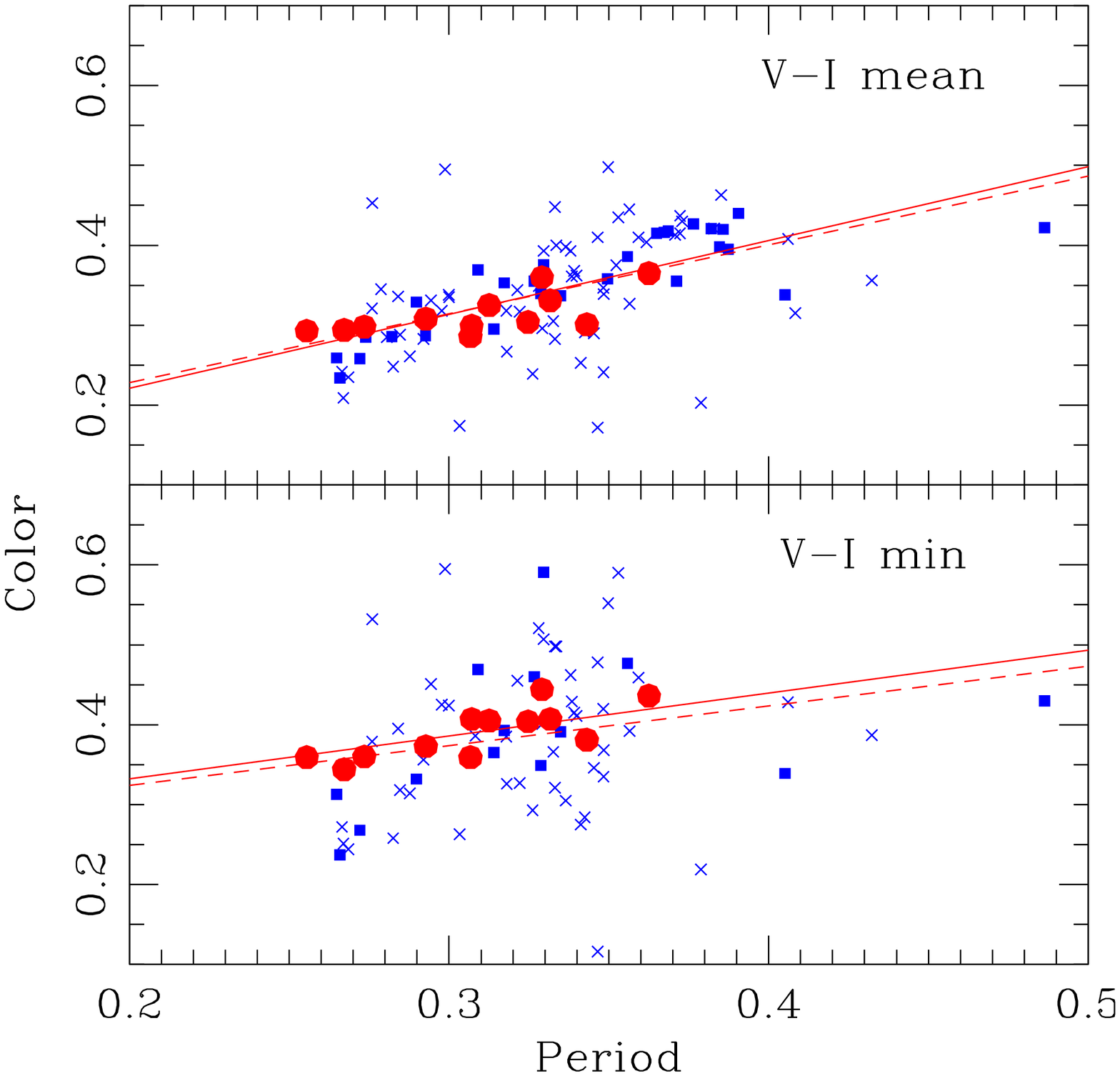}
\vskip0pt
\caption{As for Figure \ref{fig_fit1}, but regressed against pulsation
period in days.}
\label{fig_fit2}
\end{figure*}

Here we report preliminary results obtained by Tyler Anderson as part
of his 2012 M.Sc. thesis.  He focused on two definitions of color, the
intensity-mean $\langle V \rangle - \langle I \rangle$, and the color
at minimum light, $(V-I)_{min}$ using the Sturch (1966) definition of
minimum light, $0.5 \leq \phi \leq 0.8$ developed for RRab.  As shown
in the previous section, a phase range of 0.45-0.65 is more
appropriate for RRc, so the following results offer only a preliminary
look at this relation.  Tyler began by correcting each star's
observed color for the small amount of reddening estimated from the
reddening maps of Schlegel et al. (1998):
\begin{equation}
(\langle V \rangle - \langle I \rangle)_0 = (\langle V \rangle - \langle I
\rangle) - 1.24 E(B-V),
\end{equation}
with a similar definition for $(V-I)_{min,0}$, and where the color
conversion factor of 1.24 is from Cardelli et al. (1989).  He then
performed leasts-squares regressions of the form
\begin{equation}
(\langle V \rangle - \langle I \rangle)_0 = c_0 + c_1 x,
\end{equation}
where $x$ is an independent variable: the pulsation period $P$, the
stellar metallicity [Fe/H], or the light curve amplitude, $A_V$, taken
in turn.  For each regression, he computed the rms of the points
around the best fit line as a measure of goodness of fit.  The results
are shown in Table 2 and Figures \ref{fig_fit1} and \ref{fig_fit2}.

\begin{flushleft}
\begin{deluxetable*}{ccccc}
\tabletypesize{\normalsize}
\tablecaption{Preliminary Color Relations for RRc}
\tablewidth{0pt}
\tablehead{ \\ \colhead{Color}   & \colhead{$x$} &
       \colhead{$c_0$} & \colhead{$c_1$} & \colhead{rms} 
}
\startdata
$(\langle V \rangle - \langle I \rangle)_0$ & none & 0.342 & 0.000 &
0.054 \\
$(\langle V \rangle - \langle I \rangle)_0$ & $P$  & 0.036 & 0.924 &
0.031 \\
$(\langle V \rangle - \langle I \rangle)_0$ & [Fe/H] & 0.222 & --0.074
& 0.044 \\
$(\langle V \rangle - \langle I \rangle)_0$ & $A_V$ & 0.324 & 0.038 & 0.054 \\
$(V-I)_{min,0}$   & none & 0.394 & 0.000 & 0.072 \\
$(V-I)_{min,0}$   & $P$  & 0.225 & 0.536 & 0.073 \\
$(V-I)_{min,0}$   & [Fe/H] & 0.272 & --0.086 & 0.069 \\
$(V-I)_{min,0}$   & $A_V$  & 0.226 & 0.332 & 0.059 \\
\enddata
\end{deluxetable*}
\end{flushleft}

Figure \ref{fig_fit1} shows a weak correlation between color and
metallicity, with the cluster stars having a marginally steeper slope.
Figure \ref{fig_fit2} shows a stronger correlation between color and
period, again with the cluster stars having a marginally steeper
slope.  Plots of the residuals from this fit versus metallicity and
amplitude show little or no slope, indicating that period and
metallicity are themselves related.  Expanding the model to a
color-period-metallicity is thus not warranted.  Plots of color versus
amplitude, $A_V$ show little correlation.

\section{Conclusions and Future}

Our best, preliminary predictor of the intrinsic color of RRc variables is
\begin{equation}
(\langle V \rangle - \langle I \rangle)_0 = 0.036 + 0.924 P,
\end{equation}
with rms = 0.031 mag.  This scatter is only slightly larger than the
best relation for RRab stars from Guldenschuh et al. (2005),
indicating that RRc are also valuable tools for measuring interstellar
reddening.  We show that the cluster and field RRc obey approximately
the same relation, further extending the utility of RRc as a reddening
indicator.

In the summer of 2013, undergraduate student Paul Husband is working
to refine and extend these relations by adding more field and cluster
stars, particularly at the metal-poor end of the relation.  He is
revising the minimum-light color measurements using the more
appropriate phase range of 0.45-0.65, and is extending Tyler's work in
$V-I$ to $B-V$ using all three color definitions.  We will also use
the compiled data to see whether the color changes for stars
exhibiting the Blazhko effect when they are observed at different
phases in their Blazhko cycles.

Acknowledgements: P.H. and A.C.L.  acknowledge support from BGSU's
SETGO Summer Research program, which is funded by the NSF.

\end{document}